# Wedge - Shaped Absorbing Samples Look Left Handed: The Problem of Interpreting Negative Refraction, and its Solution.

M. Sanz[1], A. C. Papageorgopoulos[1], W.F. Egelhoff, Jr.[2], M. Nieto-Vesperinas[3] and N. Garcia[1].

1. Laboratorio de Física de Sistemas Pequeños y Nanotecnología. Consejo Superior de Investigaciones Científicas, Serrano 144, Madrid 28006, Spain.
2 National Institute of Standards and Technology, Gaithersburg, Maryland 20899, USA
3 Instituto de Ciencia de Materiales de Madrid. Consejo Superior de Investigaciones Científicas, Campus de Cantoblanco, Madrid 28049, Spain.

**ABSTRACT**

We report experiments of light transmissivity at wavelengths: 532 and 400 nm, through an Au film with a wedge shape. Our results mimic the negative refraction reported by others for so-called left handed materials. A mimic of negative refraction is observed, even though this medium is well known to be right handed, and thus its refractive index has a positive real part. Analogous results are obtained with a glass wedge at 320nm where absorption dominates. The experiment is explained by the wave losses that dominate over propagation, like in the observation of negative refraction, already reported in developed metamaterial wedges. We design and propose an experiment with metamaterials by using thicker wires, in correspondence with light experiments that should conclusively determine whether refraction is positive or negative.

Recently, left handed materials (LHM) (*1*) have received much attention in connection with electromagnetic wave propagation. Among other effects, they should exhibit refraction at negative angles. Nevertheless, their physics in this respect is very similar to that of metals. The proposed LHM (*2*) has $Re(\varepsilon)<0$ and $Re(\mu)<0$, which defines $Re(n)<0$, where $\varepsilon$, $\mu$ and n are the dielectric permittivity, magnetic permeability and the refraction index, respectively; *Re* denotes the real part. However, such a statement of these optical parameters is really meaningless, unless further specifications are introduced, because if $Re(\varepsilon)<0$ and $Re(\mu)<0$, the condition of positive energy implies the existence of dispersion. Hence, both the permittivity and permeability are frequency dependent and contain imaginary parts. As a consequence, the refractive index has $Im(n)>0$. In these circumstances $Re(n)$ plays no role at all since the propagation conditions are determined by the choice of the sign of $Im(n)$. This sign should be such that the wave dissipates its energy as it propagates, as brilliantly discussed in Ref. (*3*), thus $Im(n)>0$. This implies, if $Re(\varepsilon)<0$ and $Re(\mu)<0$, that $Re(n)<0$. Notice, however, that this last condition on n is imposed by the restriction $Im(n>0)$, and not viceversa. Then, if $Im(n)>0$ there are modes in the medium that do not propagate. This is precisely the case of metals. Whether propagation or attenuation dominates in such a medium, is determined by the transmissivity, which is larger or smaller than 1/e, respectively, when the wave has traversed a distance of several wavelengths. In a slab, if propagation dominates, one observes the presence of Fabry-Perot like oscillations, with decay smaller than 1/e, of either the transmissivity or the reflectance versus the frequency or the slab thickness. This is the case of transparency in dielectrics. The opposite occurs, namely no oscillations, when attenuation is dominant.

Naturally, the idea of negative refraction, coming from a LHM, has attracted the interest of scientists, because this may constitute a new area of science with technological applications. In our opinion, this is a complicated territory, although it deserves study, especially because new consequences may appear. The experimental work has been so far conducted by researchers of the University of California at San Diego (*1*). They have claimed the verification of a negative refractive index. The experiment was done in the microwave range, around $10^{10}$Hz, using a wedge shaped sample of metamaterial *by showing that the light is emitted at negative angles, that, however, correspond to the region where the thinnest part of the wedge is*!!. The point is that these metamaterials are made of metallic wires and rings, embedded in a dielectric. Certainly, mixing metallic elements with very thin wires (0.003 cm thick) and microwaves is not the best procedure to cancel the imaginary part of the refractive index. This cancellation is due to the large imaginary part of the permittivity of metals in the microwave range. In a recent paper (*4*), we have shown that the losses due to Im(n) are large enough to destroy propagation. Only decaying waves exist in the medium, and this invalidates the experimental proof of negative refractive index (*4*). Furthermore, no experiment performed so far, with the thin wire used (*5*), exhibits any oscillatory condition of the transmissivity that might indicate propagation (see references in (*1*)). All the experiments show about the variation of the transmissivity is a sharp pass band at a given frequency, with a fast decay outside this band. Certainly, this is not a signature of propagation.

In this paper we present measurements of the transmissivity of light at wavelengths $\lambda$=532 and 400nm, through an Au wedge, as well as for a glass wedge at 320nm. The thickness of the wedge is scaled to the wavelength in the same way as in the LH metamaterial wedge used (*1*). The experiment shows that the measured transmitted intensity apparently points towards negative refraction angles, namely, towards the thinner region of the wedge as in the experiment of Shelby (*1*), even if the angle of the Au wedge is only $10^{-4}$ rad. Certainly, this result has nothing to do with negative refraction, but with light attenuation in the material. The material is of course more transparent in the thinner region. Motivated by these observations, we design and propose an experiment that may determine the sign of the refraction angle, and hence of n.

An Au wedge on glass was prepared at NIST. It is 5 mm long and the thickness varies from 25 nm to 400 nm, which produces a wedge angle of $10^{-4}$ rad. Notice that, except for absorption effects, this sample has for all matters practically parallel faces. It has been built in such a way that it has a difference of approximately one wavelength $\lambda$ between the thicker and the thinner widths. This scales with the experiment of Ref.1. The light measurements were performed, in the Laboratorio de Física de Sistemas Pequeños y Nanotecnologia (CSIC), with a streak camera having a slit aperture of 6mm, which is very convenient in order to pick up the transmitted signal at once through a bunch of optical fibers. The camera is operated in focus mode just as a detector, without using its time resolution. The laser beam, whose profile is depicted in Fig.1(a), and is measured as shown in the upper frame of Fig.2, having its peak at the center of the wedge faces (red and blue color represent maximum and minimum intensity respectively). When the light impinges the wedged sample, it is maximum at its center (red) (cf. Fig. 1(a)), however, the transmitted beam is maximum at x=1 mm, (see Fig. 1(b)) as expected from the larger transparency of the Au wedge in its thinner region where absorption is smaller, and so is the associated exponential decaying behavior of the wave intensity in the metal. This is confirmed in Fig.1(c), where the value of the measured transmissivity T is plotted versus the transversal ordinate X, sowing the typical exponential decay. The agreement with the theory is excellent using the permitivity ε for Au (*6*) at two values of λ (532 and 400nm), thus proving that the sample surface is smooth compared to λ and no scattering effects exist. The sample was checked with the STM, which showed a grain size and corrugation of 40 and 10nm, respectively. Fig. 1(b) (left hand side) shows a sketch of the region where the intensity appears located in X (indicated by red color). The angle θ of the transmitted intensity peak is plotted (cf. right hand side of Fig. 1(b)) as a function of the distance d between the detector (array of optical fibers) and the sample, as well as a function of the apparent angular width θ_op of the detected beam. These two angles are plotted versus d in the right hand side of Fig. 1b, by using the data of Fig.2 as shown by the streak camera. In this figure, the panels represent the detected transmitted beam in the following conditions: no sample (top panel), and with sample at distances d from the detector: 0.5, 2.5, 10, 15 and 25mm. The left panels are a color plot of the spatial distribution of intensity detected by the camera along the X -direction, whereas the right panels are the digitized peak data of the left panels along X. As seen, the intensity is very localized at short distances d, and slightly spreads as d increases. Fig. 1(c) shows log T versus X and the corresponding thickness of Au traversed by the light. It can be observed that the exponential spatial distribution of intensity, sketched in Fig.1(a), very accurately corresponds to the measurements of Fig. 1(c).

We remark that this experiment with the Au wedge, is an accurate replica of the one performed (*1*) to prove negative refraction. The results of both experiments agree in scale according to their respective wavelengths: visible light for Au, and microwaves for the metamaterial. In the case of Au, apparent super negative refraction appears, and thus LH behavior would be inferred because the wedge angle is $10^{-4}$ Rad only. The apparent angle of refraction is even -70º when the detector is near the sample, if this angle is measured at finite distance d as in (*1*), and as illustrated in Fig. 1(b). Notice that in the metamaterials experiment the detector is located at 15cm from the sample, but this is not a far-field condition for a $\lambda \approx 3$cm. It is evident, however, that $\theta$ is not the appropriate angle of refraction, which should be considered between the surface normal and the axis of the {\it emerging beam}, and not between the surface normal and the position vector of the detection point as in (*1*) and as illustrated in Fig. 1(a). Gold is certainly not a LHM because µ=1 in the visible, *but by interpreting the refraction angle as above, it nevertheless looks like one.* As stated, this effect is due to a combination of the wedge shape of the sample and the absorption represented by the imaginary part of ε in the case of gold, and of both µ and ε in the case of the metamaterial dealt with in Ref.(*1*). Thus, this is what takes place in the experiment (*1*), as theoretically demonstrated by Garcia and Nieto-Vesperinas (*4*) (see bottom panel of Fig.2). They showed how the losses dominate wave propagation in that metamaterial, thus making the waves highly inhomogeneous and exponentially decaying. We therefore believe that, due to this absorption effect, it is impossible to determine whether the sample is left-handed or right-handed, and thus whether refraction is negative or positive, in a wedge geometry. Also we notice that Au for the frequencies at hand is highly dispersive but dispersion does not plays role in explaining the experiments. The explanation comes from an absorbing effect not from dispersion (*4,7*). Analogous result are obtained by using a wedge glass plate (5mm long and from 1 mm to 2 mm thick) with $\lambda \approx 320$nm where there is absorption. Notice that for the glass is a dielectric: *Re*(n)>0 and *Im*(n)>0, while for Au Re(n)<0 and nevertheless the transmissivity behaves equally in both cases. Therefore no matter what the material is, be metallic Au or dielectric glass, if there is absorption the transmissivity looks left handed. In the experiments so far the registered transmisstivity is $10^{-2} - 10^{-3}$ and this is absorption. So far we have discussed the experiments in mematerials using strip radius 0.003 cm thick (*5*). In this case the losses are important, however these may be reduced by increasing the radius size by a factor 10 to 20 because the losses are inversely proportional to the wire cross section (*3,8*). Therefor the experiments should be tried with thicker wires in the way proposed below.

However, there is a way to avoid these inhomogeneous absorption problems produced by a wedge, and check the sign of the refraction angle in the metamaterials, as well as in any other composite material that can be produced with magnetic elements near the ferromagnetic resonance: For this case, we have designed, and propose, the following experiment depicted as in Fig.3: It consists of measuring the beam displacement $\Delta X$ due to the refraction of the beam at both faces of a slab of parallel faces, as it traverses the medium. The incident beam impinging at an angle $\theta_1$. We have performed this experiment for a glass plate using the streak camera, and have measured the displacement $\Delta X_{exp}$, which is in excellent agreement with the theoretical displacement $\Delta X_{cal}$ (see Fig. 3, upper panel). *We propose the same kind of experiment to verify a negative refractive index,* (see the lower panel of Fig.3). The displacement provoked by negative refraction should be larger by a factor *2b.tg$\theta_2$* than the displacement due to positive refraction . For example, for b≈λ≈ 3cm and *Re*(n)=-1, the transmissivity decays to $10^{-2}$ of its incident value and we obtain that $\Delta X_{cal} \approx 10.30$cm; which is a displacement clearly observable. We believe that this is the only way to assess a LHM in the microwave region where losses dominate, as in metamaterials, and other materials that may have metallic magnetic grains or wires, immersed in a dielectric matrix. The reason is that even if the losses dominate, these are equal and homogenous in all parts of the film; i.e. the losses do not vary along the X-direction of the film. Then, what remains is the influence of *Re*(n). This is the kind of experiment to neatly experimentally verify negative refraction, if there is such a phenomenon. But the claims of observation of negative refraction by using a wedge do not fulfill this requirement, as illustrated in this work.

ACKNOWLEDGMENTS

We thank M. Muñoz, E. Ponizowskaya and C. Guerrero for valuable help. This work has been supported by the Spanish DGICyT

FIGURE CAPTIONS

Fig.1 (a) A sketch of the experimental set up. An incident beam normal to the wedge sample with maximum intensity at the center of the plate comes out with maximum intensity in the thinner region of the wedge due to absorption. The entire intensity is measured by a bunch of optical fiber focused to the streak camera. The color scale bar is for the transmitted intensity, while for the incident intensity the red corresponds to unity. (b) lhs kinematics of the problem with the angles $\theta$ and $\theta_{op}$, notice that these changed with the detector distance from the sample; rhs are the angles as measured from the intensity recorded in Fig.2. (c) Variation of the transmissivity T versus X, lower scale and the sample thickness, upper scale showing an exponential decay behavior. Circles and crosses are experiments and lines is the theory using the measured dielectric constants (*6*) showing ~~an~~ excellent agreement, and no scattering in the sample indicating a flat Au surface. This point is supported by the STM image that shows a grain size of 40 nm and a roughness of 10nm, both much smaller than the values of $\lambda$. Notice that the z scale of the roughness is magnified as indicated by the bar.

Fig.2 Streak camera measurements as the optical fiber separate from sample. The top panel is the beam picture without sample. Notice that the intensity spreads slowly as d is increased. The numbers at the side scale the intensity. On the right we show the same pictures are digitized. The intensities for the sample are normalized to the incident intensity (top panel) and the thick dots represent the position of the fiber (see Fig.1a).

Fig 3. ***The proposed experiment***. Top panel shows the experimental displacement measured in the streak camera with the parameters indicated. Lower panel is the proposed experiment for a LHM. This is a more convenient way to possibly observe negative refraction index. Notice that the displacement is much larger for the LHM.

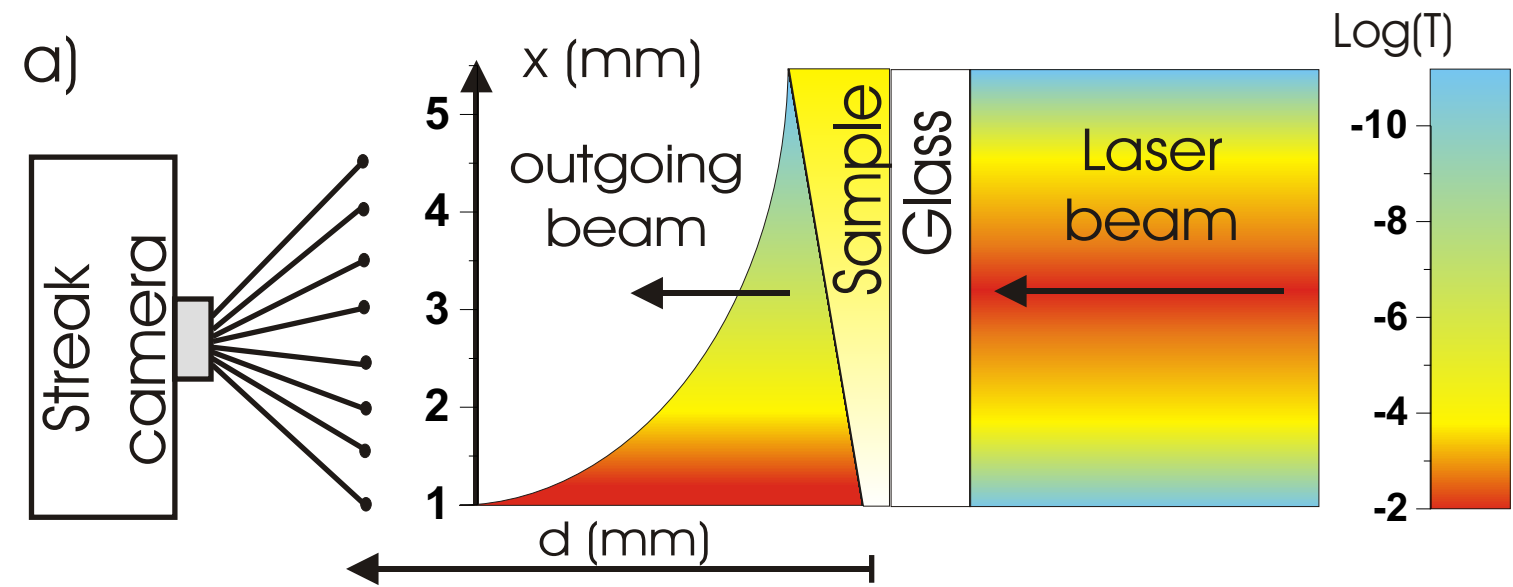
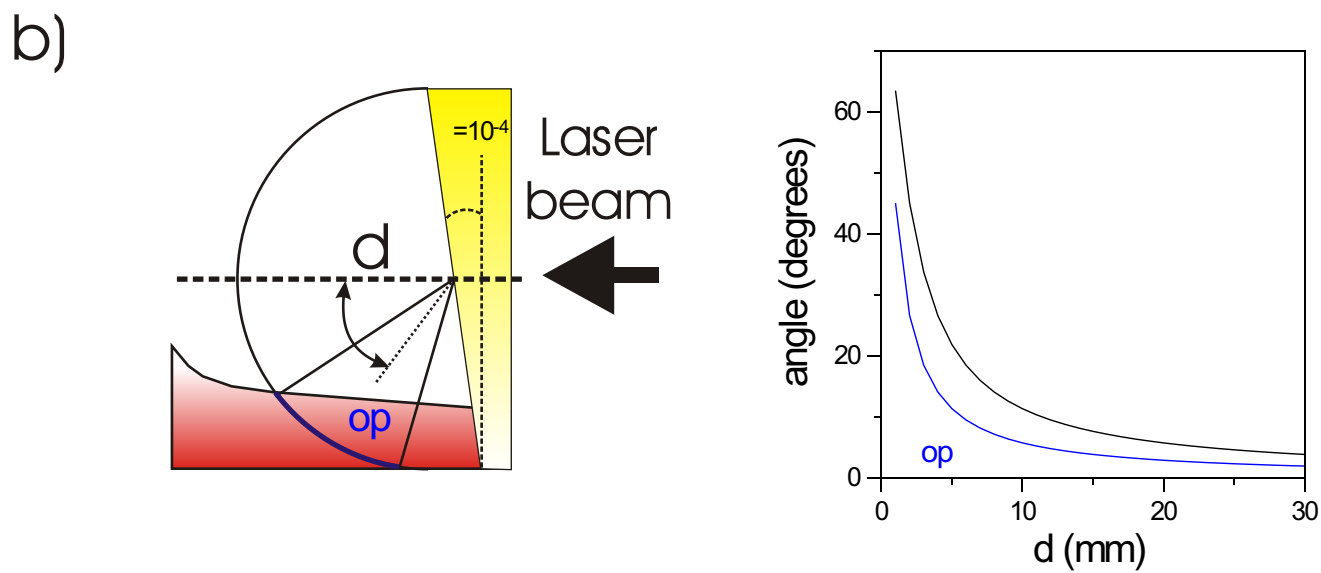
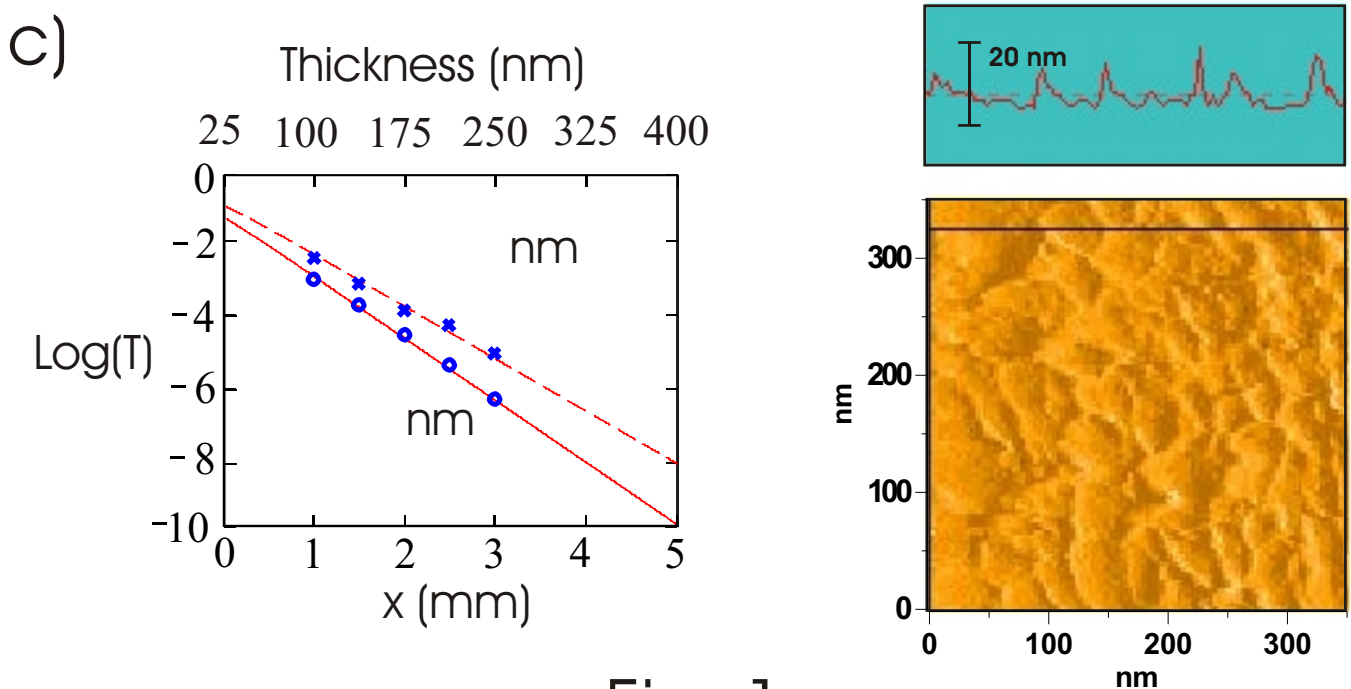

Fig. 1

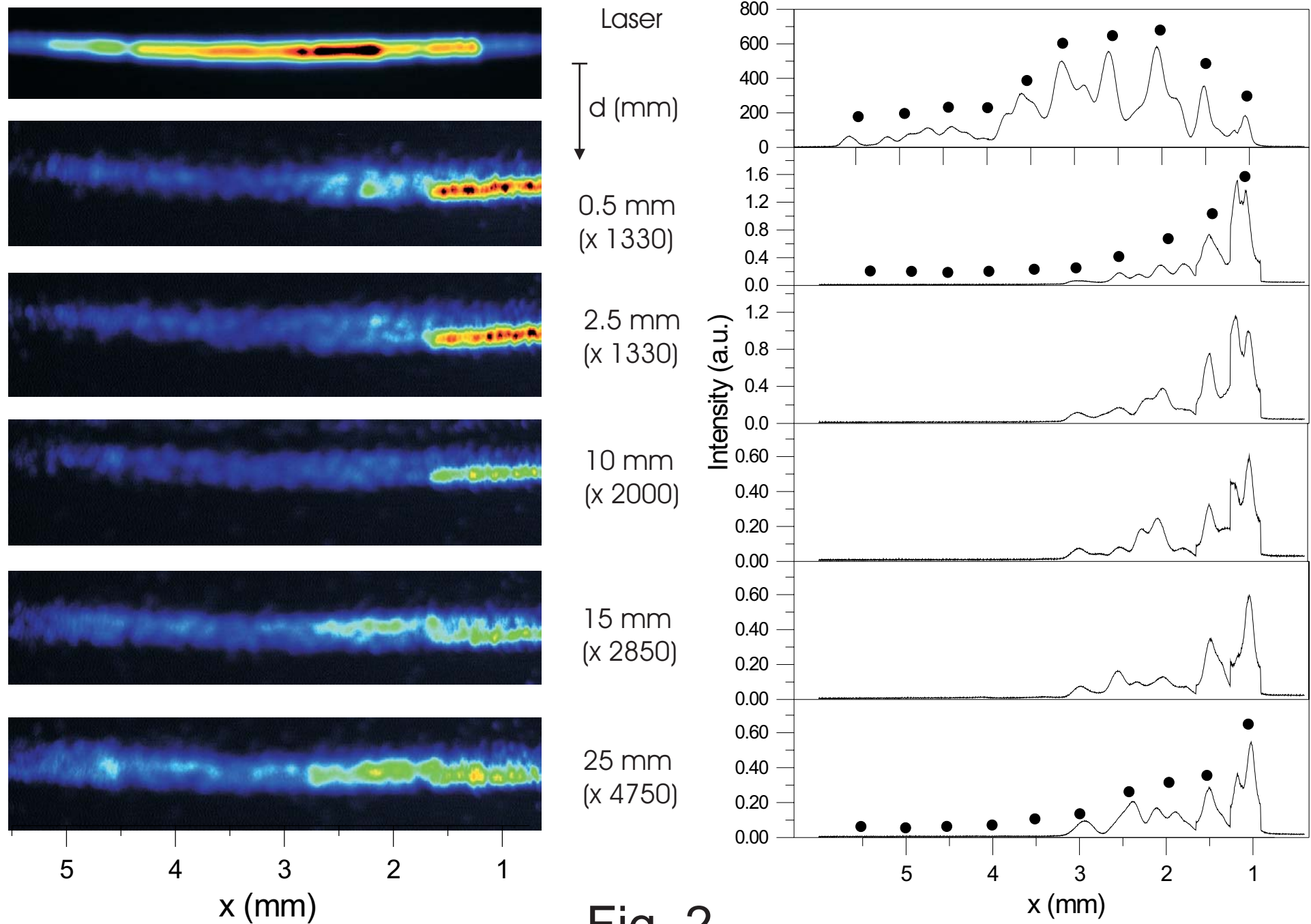

Fig. 2

**Proposed experiment**

$\theta_1 = 60°$

$n = 1.45$

$\Delta x_{cal} = b(tg\theta_1 - tg\theta_2) = 0.98\,mm$

$\Delta x_{exp} = 0.95\,mm$

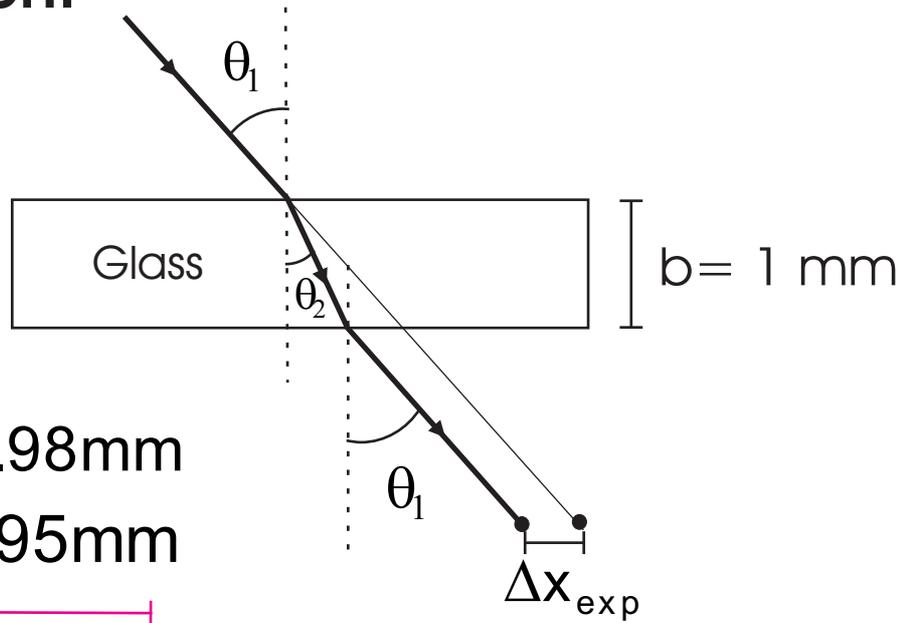

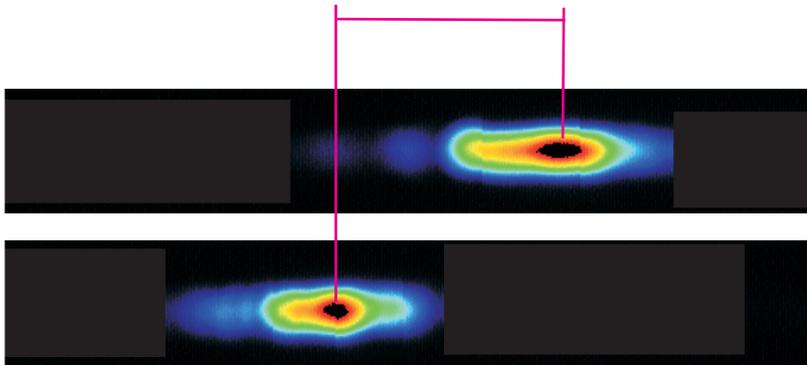

$\theta_1 = 60°$

$Re(n) = -1$

$\Delta x_{cal} = b(tg\theta_1 + tg\theta_2)$

$\Delta x_{cal} \approx 10.30\,cm$

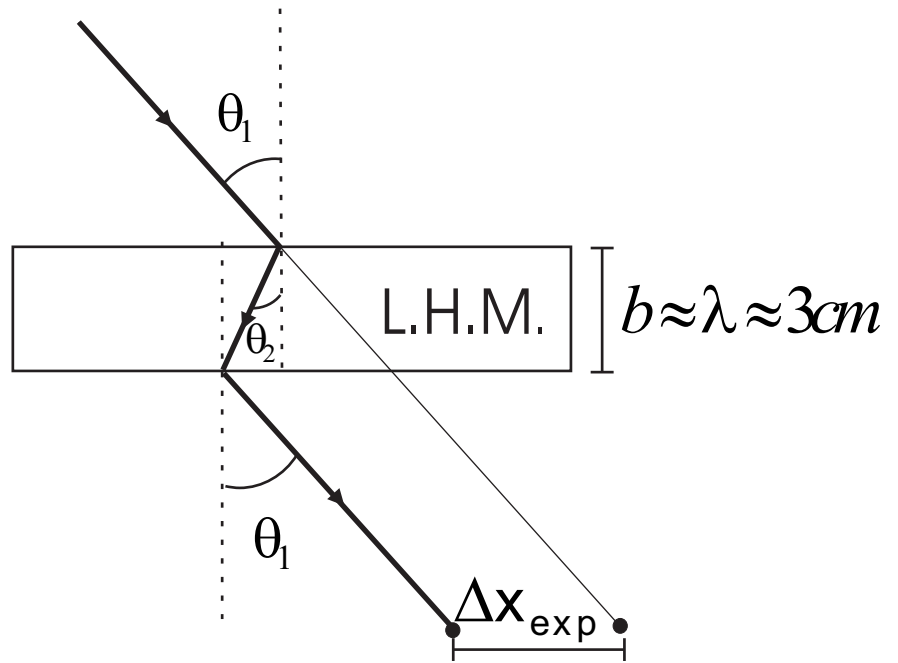

Fig. 3